\documentclass[structabstract]{aa}

\usepackage{graphicx}
\usepackage{txfonts}
\usepackage{natbib}
\bibpunct{(}{)}{;}{a}{}{,}    

\begin{document}

\title{H$_2$ infrared line emission from the ionized region \\ of planetary nebulae}

\subtitle{ }

\author{
I. Aleman
\inst{1,2}
\and
R. Gruenwald
\inst{1}
}

\offprints{I. Aleman}

\institute{
Instituto de Astronomia, Geof\'{\i}sica e Ci\^encias Atmosf\'ericas (IAG-USP), Universidade de S\~ao Paulo, Cidade Universit\'aria, Rua do Mat\~ao 1226, S\~ao Paulo, SP, Brazil, 05508-090\\
\email{isabel@astro.iag.usp.br}
         \and
Jodrell Bank Centre for Astrophysics, The Alan Turing Building, School of Physics and Astronomy, The University of Manchester, Oxford Road, Manchester, M13 9PL, UK.
}

\date{Received May 12, 2010; accepted December 23, 2010}

\abstract
   {The analysis and interpretation of the H$_2$ line emission from planetary nebulae have been done in the literature by assuming that the molecule survives only in regions where the hydrogen is neutral, as in photodissociation, neutral clumps, or shocked regions. However, there is strong observational and theoretical evidence that at least part of the H$_2$ emission is produced inside the ionized region of these objects.}
   {The aim of the present work is to calculate and analyze the infrared line emission of H$_2$ produced inside the ionized region of planetary nebulae using a one-dimensional photoionization code.}
   {The photoionization code Aangaba was improved in order to calculate the statistical population of the H$_2$ energy levels, as well as the intensity of the H$_2$ infrared emission lines in the physical conditions typical of planetary nebulae. A grid of models was obtained and the results then analyzed and compared with the observational data.}
   {We show that the contribution of the ionized region to the H$_2$ line emission can be important, particularly in the case of nebulae with high-temperature central stars. This result explains why H$_2$ emission is more frequently observed in bipolar planetary nebulae (Gatley's rule), since this kind of object typically has hotter stars. Collisional excitation plays an important role in populating the rovibrational levels of the electronic ground state of H$_2$ molecules. Radiative mechanisms are also important, particularly for the upper vibrational levels. Formation pumping can have minor effects on the line intensities produced by de-excitation from very high rotational levels, especially in dense and dusty environments. We included the effect of the H$_2$ molecule on the thermal equilibrium of the gas, concluding that, in the ionized region, H$_2$ only contributes to the thermal equilibrium in the case of a very high temperature of the central star or a high dust-to-gas ratio, mainly through collisional de-excitation.}
   {}

\titlerunning{H$_2$ IR emission from the ionized region of PNe}

\maketitle

\section{Introduction}

\defcitealias{Aleman_Gruenwald_2004}{Paper~I}

Since the first detection of H$_2$ in a planetary nebula (PN) by \citet{Treffers_etal_1976}, this molecule has been detected in many PNe both in ultraviolet (UV) absorption and infrared (IR) emission \citep[e.g.][]{Beckwith_etal_1978, Zuckerman_Gatley_1988, Aspin_etal_1993, Kastner_etal_1996, Hora_etal_1999, McCandliss_etal_2007, Sterling_etal_2005, Herald_Bianchi_2004}. The H$_2$ molecules can be excited by UV photons, collisions with the gas, or by formation on excited levels. The contribution of each mechanism to the population of the H$_2$ levels depends on the physical conditions of the gas. Up to now, analyses of the excitation mechanism of the H$_2$ molecules producing the observed IR lines have been inconclusive \citep{Dinerstein_1991, Shupe_etal_1998, Hora_etal_1999, Speck_etal_2003, Rosado_Arias_2003, Likkel_etal_2006, Matsuura_etal_2007}. 

In the literature, the H$_2$ IR emission lines from PNe is usually analyzed under the assumption that H$_2$ molecules only exist in regions where the hydrogen is neutral, such as photodissociation regions \citep[PDRs;][]{Tielens_1993, Natta_Hollenbach_1998, Vicini_etal_1999, Bernard-Salas_Tielens_2005}, shocked regions between the expanding envelope and the wind of the progenitor \citep{Gussie_Pritchet_1988, Natta_Hollenbach_1998}, or neutral clumps inside the ionized region \citep{Beckwith_etal_1978, Gussie_Pritchet_1988, Reay_etal_1988, Tielens_1993, Schild_1995, Speck_etal_2002}. 

On the other hand, there is strong observational and theoretical evidence that at least part of the H$_2$ emission is produced inside the ionized region of PNe. Some authors have noticed that the morphology of some PNe shown by images taken in the H$_2$ 1-0 S(1) infrared line is very similar to those taken in [\ion{N}{II}], [\ion{S}{II}], and [\ion{O}{I}] forbidden optical lines and in hydrogen recombination lines, which are produced in the ionized region \citep{Beckwith_etal_1978, Beckwith_etal_1980, Reay_etal_1988, Webster_etal_1988, Zuckerman_Gatley_1988, Balick_etal_1991, Schild_1995, Allen_etal_1997, Guerrero_etal_2000, Lopez_etal_2000, Arias_etal_2001, Bohigas_2001, Speck_etal_2002, Speck_etal_2003}. Excitation temperatures of approximately 1000 to 2000 K are inferred from observations of H$_2$ IR emission from some PNe \citep[][and references therein]{Hora_etal_1999, Likkel_etal_2006}. Such high excitation temperatures indicate that the molecule is excited by a strong UV radiation field, since collisions or shocks cannot explain the presence of lines from excited levels with vibrational numbers over three \citep{Hora_etal_1999, Black_vanDishoeck_1987}. Furthermore, in a previous paper \citep[hereafter \citetalias{Aleman_Gruenwald_2004}]{Aleman_Gruenwald_2004}, we calculated the density of H$_2$ inside the ionized region of PNe, showing that H$_2$ can survive in a partially ionized region with moderate temperature where neutral and ionized species coexist. Since such regions can be large when the temperature of the ionizing star is high, the ionized region can be a potential contributor to the total H$_2$ line emission in some PNe.

In the present paper we study the contribution of the ionized region to the observed H$_2$ IR line emission of PNe. The H$_2$ IR line intensities are calculated with a photoionization code. The effects of the temperature and luminosity of the central star, gas density, and dust-to-gas ratio on the line emission are studied. The models are described in Sect. \ref{mod}. In particular, the formalism adopted for calculating the H$_2$ energy level population and the intensities of the H$_2$ emission lines, as well as the included H$_2$ excitation mechanisms are described in detail. Results are discussed in Sect. \ref{res}. A summary of the conclusions and final comments are presented in Sect. \ref{final}.

\section{Models} \label{mod}

To calculate the intensity of H$_2$ infrared lines emitted by the studied region, the molecular density and the population of each rovibrational energy level of the electronic ground state must be known. Both the molecular density and the level population depend on the location inside the nebula, since the radiation field, and thus the physical conditions of the gas, depends on the distance to the ionizing source. The physical conditions of the nebular gas were obtained with the one-dimensional photoionization code Aangaba \citep{Gruenwald_Viegas_1992}. For a description of the code and of the calculation of the H$_2$ density see \citetalias{Aleman_Gruenwald_2004}. For the present study, the calculation of the H$_2$ level population and line intensities were introduced into the numerical code. The formalism adopted for these calculations is described below.

\subsection{The H$_2$ energy level population}

In Aangaba, the population of each H$_2$ level can be calculated by assuming either local thermodynamical or statistical equilibrium. In the first case, the level population is set by the Boltzmann distribution and only depends on the gas temperature. We assume statistical equilibrium, where the population of the energy levels is given by the balance between the rates of population and depopulation for each level. This is equivalent to stating that there is no net variation in the population of each level, i.e.,

\begin{equation}
      \frac{dn_w(v,J)}{dt} = 0 .
\label{eqest}
\end{equation}
In the equation above, $n_w(v,J) $ is the density of H$_2$ in an energy level with vibrational and rotational quantum numbers, respectively, $v$ and $J$ of the electronic state $w$. There is an equation as Eq. \ref{eqest} for each level. This set of equations is linearly dependent and, to solve it, we replace one of the equations by the equation of conservation 

\begin{equation}
      \sum_{v,J} n_w(v,J) = n(\mathrm{H}_2)
\end{equation}
where $n($H$_2)$ is the density of molecular hydrogen.

The expression on the left hand side of Eq. \ref{eqest} is evaluated by the algebraic sum of the population and depopulation rates of the level $(v,J)$ owing to the various mechanisms. To calculate the population in the rovibrational levels of the ground state, we included several excitation and de-excitation mechanisms (radiative and collisional), as well as the possibility that H$_2$ is produced or destroyed at a given level by chemical reactions. Equation \ref{eqest} can be rewritten as

\begin{equation}
      \frac{dn_w(v,J)}{dt}\bigg |_\mathrm{rad} + \frac{dn_w(v,J)}{dt}\bigg |_\mathrm{col} + \frac{dn_w(v,J)}{dt}\bigg |_\mathrm{chem}= 0
\label{colradfor}
\end{equation}
where the terms on the left hand side of the equation are the terms that correspond to level population, respectively, by radiative transitions, collisional transitions, and formation or destruction processes by chemical reactions. The processes included in the calculation are described in Sects. \ref{rad} to \ref{form}. To calculate the population of the upper electronic states, only radiative electronic transitions between the excited state and the ground electronic state are taken into account (see justification in Sect. \ref{col}).

The set of equations can be solved for the density of each level if the rate coefficient for each process and the densities of the reactant species are known. The total H$_2$ density, as well as the densities of H$^0$, H$^+$, H$^-$, H$_2^+$, and H$_3^+$, is given by the solution of the chemical equilibrium equations \citepalias[see][]{Aleman_Gruenwald_2004}. In the following we discuss the included mechanisms of populating and depopulating the energy levels of the H$_2$ molecule.

\subsubsection{Radiative transitions} \label{rad}

The population of the H$_2$ rovibrational levels of the electronic ground state by radiative mechanisms occurs through two main routes: electric quadrupole transitions between the rovibrational levels, involving IR photons, or electric dipole transitions to upper electronic states with posterior decay to the ground state, involving UV photons.

The transitions between two energy levels with different rotational and/or vibrational quantum numbers within the same electronic state are called rovibrational transitions. For the electronic ground state of H$_2$, X$^1\Sigma _\mathrm{g}^+$ (hereafter only referred to as X), they occur through electric quadrupole transitions, since transitions by electric and magnetic dipole are forbidden. Quadrupole transitions have much lower probabilities than the electric dipole ones. For the X state, we included 289 rovibrational levels distributed over 15 vibrational levels. The H$_2$ rovibrational transition probabilities were calculated by \citet{Wolniewicz_etal_1998}, and the energies of the levels were kindly provided by E. Roueff (2005, private communication). The selection rules for these transitions imply that the change in the rotation quantum number must be $\Delta J = 0, \pm 2$. The lines produced by rovibrational transitions of H$_2$ are in the 0.28 $\mu$m to 6.2 mm range of the electromagnetic spectrum.

Transitions where the electronic state of the molecule changes, with or without change in the rotational or vibrational quantum numbers, are called electronic transitions. Electronic transitions are allowed by electric dipole and are thus very likely to occur. They are efficient in exciting and destroying H$_2$ molecules and must then be included in the calculation of the statistical population of level X. We included electronic transitions between the ground state and the excited electronic levels B$^1\Sigma _\mathrm{g}^+$ and C$^1\Pi _\mathrm{u}$ (Lyman and Werner bands, respectively). The levels B$^1\Sigma _\mathrm{g}^+$ and C$^1\Pi _\mathrm{u}$ are referred to hereafter just as B and C, respectively. The $\Lambda$ doubling splits the energy levels of C state in two, denoted here by C$^-$ and C$^+$. For the B electronic level, 408 rovibrational levels, distributed over 38 vibrational levels, are taken into account; for C$^-$ and C$^+$, 143 and 142 rovibrational levels, respectively, are included, distributed over 14 vibrational levels. Excitation to B and C levels can be followed by decay back to the ground state on any rovibrational level (this route is known as UV pumping) or to the vibrational continuum of X leading to the molecule dissociation. This last route is called photodissociation in two steps or Solomon process, and it is described in Sect. \ref{form}. The Einstein coefficients and the wave number of the electronic transitions have been calculated by \citet{Abgrall_etal_1994} and the photodissociation fractions by \citet{Abgrall_etal_2000}. The selection rules for electronic transitions requires that the change in the rotation quantum number must be $\Delta J = \pm 1$ for Lyman band transitions, while in the case of Werner band transitions $\Delta J = 0$ for X-C$^-$ transitions and $\Delta J = \pm 1$ for X-C$^+$ transitions. The electronic excitation of the Lyman and Werner bands requires photons with energy between 6.7 and 15 eV.

\subsubsection{Collisional transitions} \label{col}

Collisions may also change the H$_2$ energy level. In photoionized regions the average energy of the particles is less than 1.7 eV, and in regions where the H$_2$ density is significant the average energy is even less. Since high energies are needed for electronic transitions ($>$ 6.7 eV), collisional electronic excitation is unlikely to occur. On the other hand, collisional rovibrational transitions may be very important in astrophysical environments and are so taken into account. We included collisions of H$_2$ with the main components of the gas, i.e., H, H$^+$, He, H$_2$, and electrons. Collisions may change the total nuclear spin by inducing a temporary non-zero dipole moment in H$_2$ (the so called reactive collisions, in opposition to the non-reactive collisions, in which there is no change in the molecular nuclear spin). 

Rate coefficients for the de-excitation of H$_2$ by collisions with H atoms were calculated by \citet{Martin_Mandy_1995}, using the quasiclassical trajectory method, for all H$_2$ rovibrational levels of the ground state, for both reactive and non-reactive collisions. \citet{LeBourlot_etal_1999} give a compilation of references for the rate coefficient of non-reactive collisional de-excitation of H$_2$ by H, He, and H$_2$ calculated by fully quantum methods, but only for H$_2$ vibrational levels up to $v =$ 3. This method of calculation is more accurate than the quasiclassical trajectory method (D. Flower 2005, private communication) and was then preferred when available.

Rate coefficients for de-excitation of H$_2$ by collisions with H$^+$ were calculated by \citet{Gerlich_1990} for pure rotational transitions of $v = $ 0, with $J_{upper}$ up to 9 and $\Delta J$ up to 9. The remaining pure rotational transitions of $v = $ 0 ($J_{upper}$ from 10 to 29) were estimated by extrapolating the existing coefficients. To approximately reproduce the behavior of the available coefficients, we assume the following extrapolations:

\begin{itemize}
\item For $J_{lower} \leq$ 7, we assume the same coefficient of the transition from the highest $J_{upper}$ of the same parity calculated by \citet{Gerlich_1990}, going to the same lower $J$ level of the transition in question. For example, the coefficient for the transition from $J_{upper} =$ 11 to $J_{lower} =$ 2 is equal to the coefficient for the transition from $J_{upper} =$ 9 to $J_{lower} =$ 2.
\item For transitions with $J_{lower} \geq$ 8, we assume the same coefficient of the highest transition calculated by \citet{Gerlich_1990} involving upper and lower $J$ levels of the same parity. For example, the coefficient for the transition from $J_{upper} =$ 14 to $J_{lower} =$ 9 is equal to the coefficient for the transition from $J_{upper} =$ 8 to $J_{lower} =$ 7.
\end{itemize}
 
For pure rotational transitions of levels $v \geq $1, we assume the same coefficients as the corresponding transition of $v =$ 0. Coefficient rates for H$_2$ vibrational de-excitation by electrons for $\Delta v$ up to 3 are given by \citet{Draine_etal_1983}. Since they do not provide the $J$-resolved rate coefficients, we assume that the transition rate for each rovibrational transition is proportional to the Einstein coefficient of the corresponding radiative transition. Only non-reactive transitions are included ($\Delta J =$ 0, $\pm$2).

In the present work we assume that the coefficients for the collisional excitation of H$_2$ by electrons are the same as those for collisions with H$^+$, because of the lack of rate coefficients for this mechanism.

\subsubsection{Formation and destruction of H$_2$ at a given energy level} \label{form}

In \citetalias{Aleman_Gruenwald_2004} the main processes of formation and destruction of H$_2$ due to chemical reactions in the ionized region of PNe were determined. Chemical reactions form or destroy H$_2$ in a given level. Since the H$_2$ molecule can be significantly produced or destroyed by these mechanisms, they must be included in the set of equations of statistical equilibrium. Unfortunately, level-resolved rate coefficients for most of the reactions are still poorly known and some assumptions must be made. The processes and our assumptions are discussed in the following paragraphs.

\paragraph{Photodissociation --}

There are two main routes for H$_2$ photodissociation: the direct and the two-step processes. In the direct process, the H$_2$ molecule is excited from the ground state to the vibrational continuum of an upper electronic state by an UV photon. The cross section for this process was calculated by \citet{Allison_Dalgarno_1969} for each of the 15 bound vibrational levels of the ground state of H$_2$. However, since they provide no information on the cross section for each rotational level, we assume that the coefficient rate is the same for all rotational levels of the same vibrational level.

As said above, the H$_2$ photodissociation in two steps begins with the excitation of the molecule from the ground electronic state to a bound rovibrational level of an upper electronic state. The subsequent decay to the vibrational continuum of the ground electronic level leads to the dissociation of the molecule. According to \citet{Stecher_Williams_1967} and \citet{Abgrall_etal_1997}, this happens in approximately 11\% of the excitations to the B and C levels. This process is particularly important because molecules can be dissociated by photons with energies smaller than the H ionization potential. As a result, the rate of this process is not significantly affected by the H column density, but depends mostly on the geometrical dilution of the UV radiation and on the H$_2$ column density (i.e., depends on the self-shielding). The Einstein coefficients for the electronic excitations and the fraction of the molecules that dissociates (dissociation fraction) have been obtained by \citet{Abgrall_etal_1994, Abgrall_etal_2000}, respectively. The effect of the H$_2$ self-shielding on the photodissociation in two steps was included, using the formalism of \citet{Black_vanDishoeck_1987}.

\paragraph{Photoionization --}

The threshold energy for the photoionization of H$_2$ is $E_\mathrm{th}(0,0) = $ 15.4 eV for the ground rovibrational level of X and is smaller for upper rovibrational levels. It is assumed here that all H$_2^+$ molecular ions are formed at the ground level. For H$_2$ in vibrational levels $v > 4$, the threshold energy of photoionization is lower than the photoionization potential of H (13.6 eV), but since the population of such levels are often very small, particularly in the recombination zone where the H$_2$ density is significant, this is not an important depopulation process. An accurate cross section for the photoionization from the vibrational ground level of H$_2$ is given by \citet{Yan_etal_1998, Yan_etal_2001}\footnote{There is a small correction to their fitting formula for the first energy interval: the coefficient of the second term should be 197.25 (H. Sadeghpour 2010, private communication).}. We assume the same shape for the cross section for the upper vibrational levels, except that it is displaced in energy by the difference in the photoionization threshold energies of the levels. Since there are no $J$-resolved cross sections, it is further assumed that the photoionization rate of a given vibrational level is distributed uniformly among its rotational levels.

\paragraph{Formation of H$_2$ on grain surfaces --}

In \citetalias{Aleman_Gruenwald_2004}, we described the model adopted for the grain surface reaction and the expression used for its rate coefficient. The coefficients for H$_2$ formation by this reaction for each energy level have been discussed in the literature \citep[and references therein]{Takahashi_Uehara_2001,Tine_etal_2003}, but they are not well established yet. We assumed the same level distribution of H$_2$ produced by this process as adopted by \citet{Black_Dalgarno_1976}.

\paragraph{Associative detachment reaction --}

The associative detachment, 

\begin{equation}
  H + H^- \to H_2 + h\nu ,
\end{equation}
is a major route for the formation of H$_2$ in the ionized region of PNe \citep[\citetalias{Aleman_Gruenwald_2004}]{Black_1978}. The rate coefficients for the formation of H$_2$ at each bound level of the ground state by this reaction were calculated by \citet{Launay_etal_1991}.

\paragraph{Charge exchange reactions --}

The H$_2$ molecule can be produced and destroyed by charge exchange reactions
\begin{equation}
  H + H_2^+ \to H^+ + H_2
\end{equation}
\begin{equation}
  H^+ + H_2 \to H + H_2^+ .
\end{equation}
The rate coefficients for these processes are given by \citet{Karpas_etal_1979} and \citet{Galli_Palla_1998}, respectively. For the first process we assume that the level distribution of the resulting molecules has the same distribution as the one used by \citet{Black_Dalgarno_1976} for molecules formed on grain surfaces, but with the appropriate energy of formation \citep[1.83 eV,][]{Hollenbach_McKee_1979}. For the destruction route we adopted the formula for the coefficients given by \citet{Galli_Palla_1998}. Since the level-resolved coefficient rates are unknown\footnote{After the conclusion of this work we found that \citet{Savin_etal_2004} had published coefficient rates for the H$_2$ destruction in each vibrational level by this process. However, as we show in the following, this reaction (and its reverse) is not an important mechanism of H$_2$ excitation, so our conclusions will not be affected by using these new coefficients}, we assume that H$_2$ molecules can be destroyed by charge exchange reaction in any rovibrational level equally.

\paragraph{Collisional dissociation --}

For the collisional dissociation of H$_2$ by H atoms we used the formula given by \citet{Shapiro_Kang_1987} for the rate coefficient, while the coefficient is that of \citet{Donahue_Shull_1991} for the dissociation by electrons. We assume that the H$_2$ molecule is dissociated following the level distribution suggested by \citet{Lepp_Shull_1983},
 
\begin{equation}
  k(v,J)=k_\mathrm{TOTAL} \frac{E(v,J)}{\sum_{(v,J)}^{} {E(v,J)}}
\end{equation}
where $k_\mathrm{TOTAL}$ is the total coefficient rate of collisional dissociation of H$_2$, and $E(v,J)$ is the energy of the level $(v,J)$.

\subsection{H$_2$ and the thermal equilibrium}

Thermal equilibrium is assumed for the calculation of the gas temperature, that is, the total input of energy in the gas per unit time and volume is balanced by the total loss of energy per unit volume. Besides the processes of gas heating and cooling associated with the atomic species and with the grains \citep[for details of how dust is included in the calculations, see][]{Gruenwald_etal_2011}, in the present work, we also included the H$_2$ molecule contribution to the energy balance of the gas. The mechanisms of gas heating due to H$_2$ are collisional de-excitation, photoionization, photodissociation, formation on grain surfaces, associative detachment, and charge exchange reaction; the mechanisms of energy loss are collisional excitation, charge exchange reaction, and collisional dissociation.

\subsection{H$_2$ infrared emission line intensities}

The emissivity of a line produced by de-excitation from a given level is the product of the population of that level by the Einstein coefficient of the transition and by the energy of the emitted photon. The emissivities of the H$_2$ IR lines are calculated for each point of the nebula. In our models, we assume that the nebula is radiation-bounded. The intensities are calculated by integrating the emissivity over the nebular volume. It is also possible to obtain solutions for matter-bounded PNe by integrating the emissivity up to a given external radius. The gas is assumed to be optically thin for the H$_2$ lines.

\subsection{Input parameters}

A grid of theoretical models was obtained for incident radiation spectrum, gas density, and dust-to-gas typical of PNe. The central star is assumed to emit as a blackbody. The gas density and the elemental abundances are assumed constant along the nebula. We use the mean values for PNe obtained by \citet{Kingsburgh_Barlow_1994} for the abundances of He, C, N, O, Ne, S, and Ar. For Mg, Si, Cl, and Fe, whose abundances are not given by \citet{Kingsburgh_Barlow_1994}, the values are those adopted by \citet{Stasinska_Tylenda_1986}, in order to roughly correct for grain depletion. Since the effect of these elements is not very important for the gas cooling, their exact proportion in the form of grains is not important for our results. Amorphous carbon grains with 0.1$\mu$m radius are assumed. 

We define a standard PN with a given set of input parameters to explore their effect on the level population and on the intensity of the lines emitted by the H$_2$ molecule. For this, we vary one of the parameters within its typical range, while keeping the other parameters fixed. Unless otherwise noted, the parameters of the models are those of the standard PN. The input parameters for the standard PN are the following: $T_{\star} = 150\,000$ K, $L_{\star} = 3000 L_{\sun}$, $n_\mathrm{H} = 10^3$ cm$^{-3}$, and the assumed dust-to-gas ratio is $M_\mathrm{d}/M_\mathrm{g} = 10^{-3}$. The temperature chosen for the central star of the standard PN is above the average for observed PNe ($\sim$ 80\,000~K) in order to highlight the effects of H$_2$.

\section{Results} \label{res}

\subsection{The transition zone}

\begin{figure}
\centering
\includegraphics[width=10.2cm]{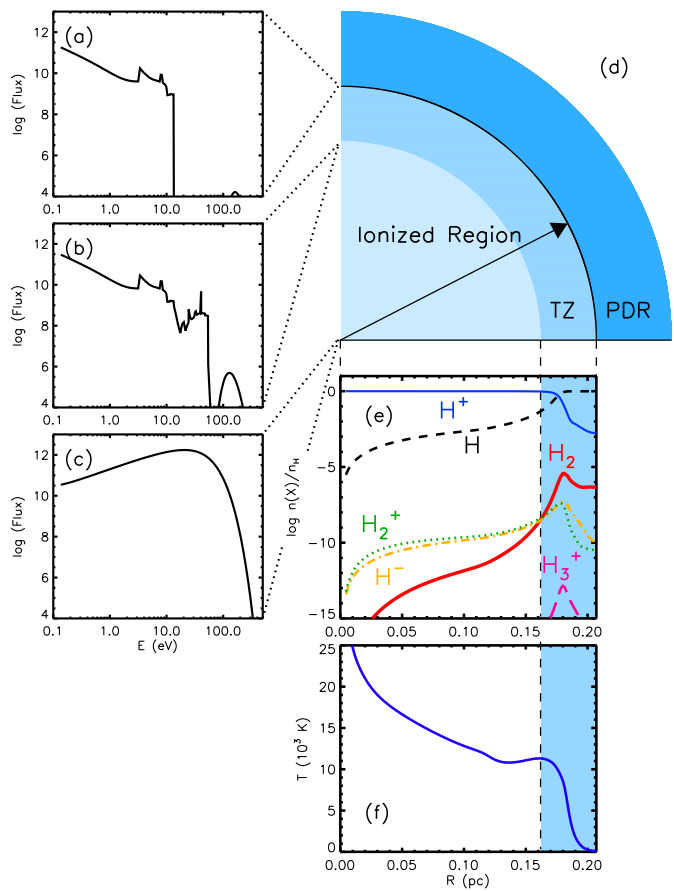}
\caption{The standard PN model. (a)-(c) Ionizing continuum spectra in three different positions of the nebula, as indicated in the cartoon representing the ionization structure of the standard PN in panel (d) (the PDR is not to scale). (e) Radial density profile of H$^0$, H$^+$, H$^-$, H$_2$, H$_2^+$, and H$_3^+$ in the ionized region. Densities are relative to the total density of H nuclei. (f) Radial gas temperature profile in the ionized region. The blue band in both panels indicates the TZ. A color version of this figure is available online.}
\label{Density}
\end{figure}

In our discussion we formally define the outer boundary of the ionized region as the location where one of the following conditions is reached\footnote{We note that the same criteria were assumed in \citetalias{Aleman_Gruenwald_2004} models, although, due to an oversight, only the ionization degree was mentioned in that paper.}: a) the fractional abundance of ionized hydrogen reaches 0.01\% or b) the gas temperature is less than 100~K. The ionization degree value is chosen following \citet{Tielens_2005}. Since for some PNe models this ionization degree is only reached in very low gas temperatures, we include the additional temperature stop criterion. In most of our models, the latter condition is reached first. We also define the transition zone (TZ) as the region between the location where the fraction of ionized hydrogen is 95\% and the outer boundary of the ionized region. A cartoon representing the ionization structure of a PN is shown in Figure \ref{Density}d.

Figure \ref{Density} shows results for the standard PN. The ionizing continuum spectrum is given for three different positions in panels (a) to (c). Panel (c) shows the incident continuum spectrum at the inner edge of the NP. The spectrum is the geometrically diluted stellar blackbody. Panel (b) shows the continuum spectra at the inner edge of the TZ. The spectrum shows absorption mainly due to hydrogen and helium, as well as diffuse UV emission features, but there is a significant flux of UV and soft X-ray photons. These high-energy photons emitted by the central star are responsible for the formation of the TZ and are very important for the physics and chemistry in this zone. Panel (a) shows the continuum spectrum emerging from the ionized region. The flux of photons with energy above 13.6~eV is significantly absorbed, indicating the inner boundary of the PDR (see Tielens (2005).

The radial distribution of H$_2$ density (relative to the total H nuclei density) for the standard PN model is shown in Fig. \ref{Density}e. In the figure, R is the distance to the central star. Relative densities of H$^0$, H$^+$, H$^-$, H$_2^+$, and H$_3^+$ are also included in this figure. The gas temperature profile is shown in Fig. \ref{Density}f. The TZ for this model is the region indicated by the blue band in Figs. \ref{Density}e and f. From these figures, it can be noticed that the relative density of H$_2$ molecules reaches a maximum in this warm and partially ionized region, as shown in \citetalias{Aleman_Gruenwald_2004}. For typical PNe models, the maximum relative H$_2$ density in the ionized region is around 10$^{-5}$, with the exception of models with hot central stars ($T_{\star} \geq$ 250\,000 K) where the density can be orders of magnitude higher. After a slight decrease, the rising of the H$_2$ density towards the more neutral zone depends on the grain density \citepalias[see][]{Aleman_Gruenwald_2004}, since the formation of H$_2$ on grain surfaces is the most important processes in this outer zone of the ionized region.

High-energy photons ($h\nu > $ 100 eV) emitted by the central star are responsible for the formation of the TZ. They can penetrate deep into the nebulae, producing this extended region of partially ionized and warm gas. Such physical conditions favor the formation and survival of H$_2$, as showed in \citetalias{Aleman_Gruenwald_2004}. The formation rates of H$_2$ molecules due to associative detachment of H and H$^-$ and to charge exchange reactions between H and H$_2^+$ are enhanced in a mild temperature gas where neutral and ionized species coexist. Furthermore, the destruction is not very effective because of self-shielding and shielding provided by atoms and grains present internally. Photons in the 13.6 to 100 eV range are absorbed by H and He atoms in the inner regions and do not contribute to the destruction of H$_2$ molecules in this region. The amount of H$_2$ inside the ionized region of PNe is then very sensitive to the thickness of the TZ.

\begin{figure}
\centering
\resizebox{\hsize}{!}{\includegraphics{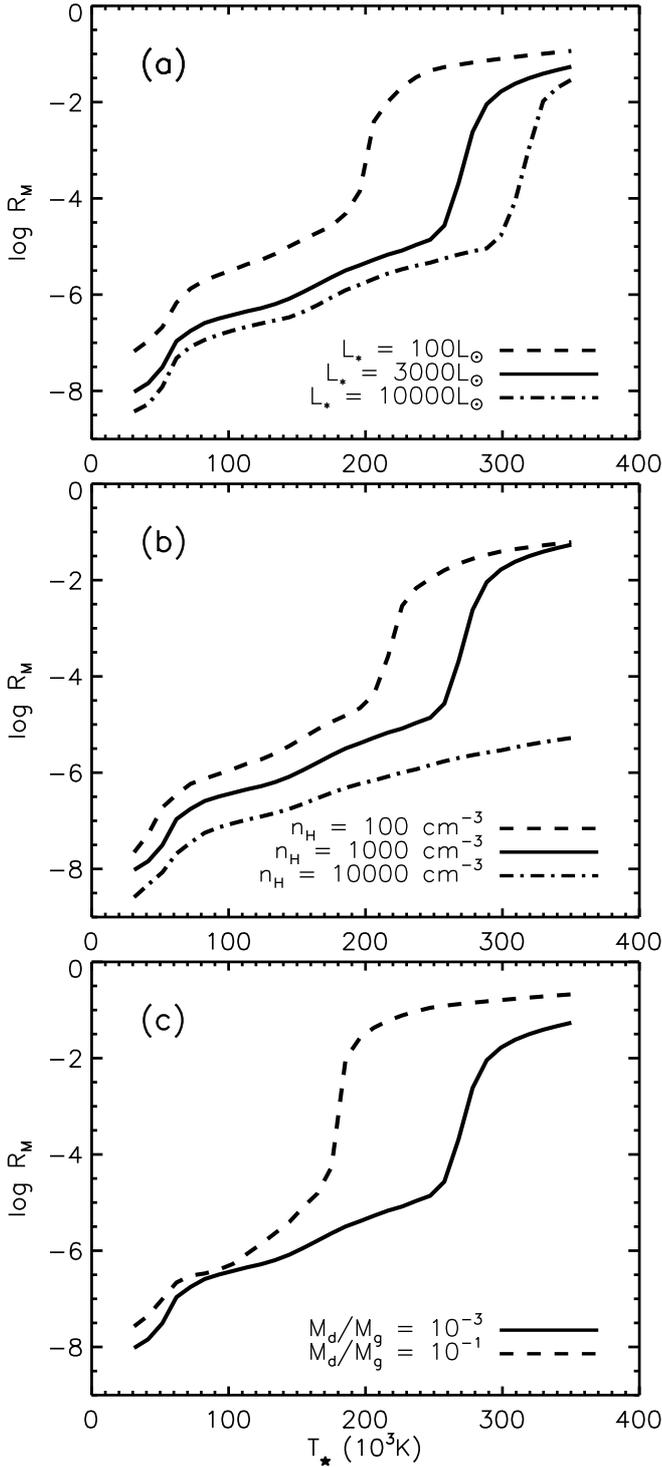}}
\caption{H$_2$ to total H mass ratio ($R_M$) as a function of the stellar temperature. Curves for different (a) $L_{\star}$, (b) $n_H$, and (c) $M_\mathrm{d}/M_\mathrm{g}$ are shown. The standard PN values are adopted for the parameters not mentioned.}
\label{rm}
\end{figure}

The available number of high-energy photons increases strongly with $T_{\star}$. As a consequence, the amount of H$_2$ in the ionized region of PNe also increases, as can be seen in Fig. \ref{rm}, where the H$_2$ to total H mass ratio ($R_M$) is shown as a function of $T_{\star}$. Curves for different values of $L_{\star}$, $n_\mathrm{H}$, and $M_\mathrm{d}/M_\mathrm{g}$ are plotted, respectively, in Figs. 2a, 2b, and 2c.

The size of the TZ does not depend significantly on the luminosity of the central star. However, a luminous central star produces more photons and is more effective in photoionizing and photodissociating molecular hydrogen. Consequently $R_M$ decreases for models with higher $L_{\star}$ (Fig. \ref{rm}a).

The value of $R_M$ also increases significantly with decreasing total H nuclei density ($n_\mathrm{H}$; see Fig. \ref{rm}b). This effect is stronger in models with hotter central stars. The stronger absorption of the radiation in the gas in models with high density produces thinner TZs. It is important to notice that we are also increasing the dust density when we increase the gas density in a model, since the dust-to-gas ratio is fixed in any given model. The effect of changing the dust-to-gas ratio ($M_\mathrm{d}/M_\mathrm{g}$) on the size of the TZ is complex since dust modifies the temperature of the gas, absorbs the UV radiation, and acts as a catalyst for the molecular formation. As a result, the size of the TZ increases with $M_\mathrm{d}/M_\mathrm{g}$ for low-temperature central stars, while the opposite effect occurs with hot stars. However, $R_M$ increases when $M_\mathrm{d}/M_\mathrm{g}$ is increased in the models, as can be seen in Fig. \ref{rm}c. This is mostly due to the increase in the rate of the reaction of H$_2$ formation on grain surfaces. When significant density of grains is available (i.e., $M_\mathrm{d}/M_\mathrm{g}>10^{-3}$), the reaction of H$_2$ formation on grain surfaces is an important process of H$_2$ production in the outer part of the TZ. 

Self-shielding is included in the present work (Sect. 2), resulting in an enhancement in $R_M$ in comparison with the models of \citetalias{Aleman_Gruenwald_2004}, particularly for models with high central star temperature, low central star luminosity, low gas density, and/or high $M_\mathrm{d}/M_\mathrm{g}$ (compare Fig. \ref{rm} with their Fig. 4).

\subsection{The effect of H$_2$ on the thermal equilibrium} \label{term}

Atomic species dominate the heating and cooling of the gas in the ionized region of PNe. If grains are present, they contribute significantly to the gas temperature (heating the gas) in a narrow zone near the inner border of the nebula and in the TZ. The increase in the temperature may be several thousand Kelvin in the first case and of a few hundred in the second.

The effect of H$_2$ on the thermal equilibrium is insignificant in most models, except in the cases of very high central star temperature and/or dust density. In these models, the net effect of the molecule is the cooling of the gas in the outer zone of the TZ, primarily through collisional excitation. For PNe with $T_{\star} =$ 350\,000~K, for example, the gas temperature is decreased by up to 5000~K, while for the standard PN the temperature structure is not altered by the presence of H$_2$.

\subsection{The H$_2$ 1-0 S(1) line emission}

The 1-0 S(1) line is produced by the radiative transition from the level $(v,J) =$ (1,3) to the level (0,1) of the ground electronic state and is one of the most intense lines of H$_2$. Its wavelength is 2.122 $\mu$m, which is in the K atmospheric window, so it can be detected by both ground- and space-based telescopes. This line has been detected in many PNe \citep[for example][]{Hora_etal_1999}.

\begin{figure}
\centering
\resizebox{\hsize}{!}{\includegraphics{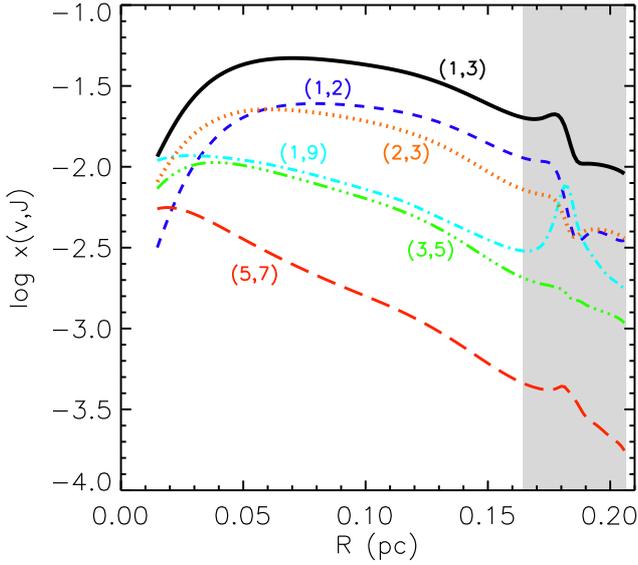}}
\caption{Relative population of some H$_2$ rovibrational energy levels of the ground electronic state as a function of the distance to the central star for the standard PN model. Levels are indicated by their vibrational and rotation numbers $(v,J)$. The gray band indicates the TZ. A color version of this figure is available online.}
\label{Pov}
\end{figure}

\begin{figure}
\centering
\resizebox{\hsize}{!}{\includegraphics{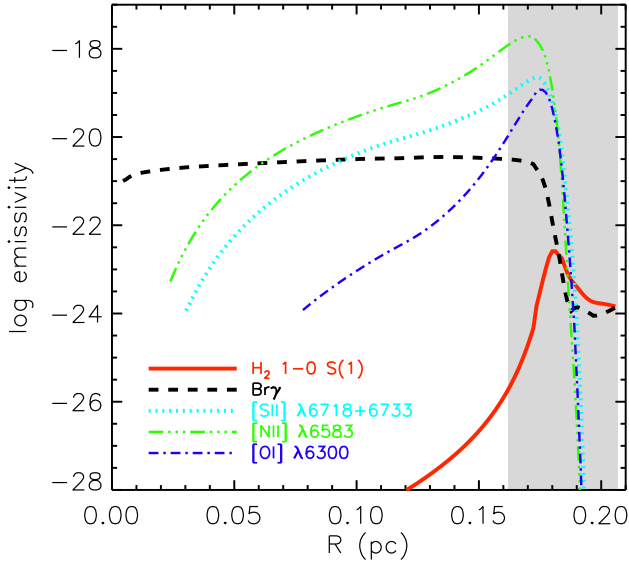}}
\caption{Emissivities (in erg cm$^{-3}$ s$^{-1}$) of the lines H$_2$ 1-0 S(1), Br$\gamma$, [\ion{S}{II}], [\ion{N}{II}], and [\ion{O}{I}] as a function of the distance to the central star for the standard PN model. The gray band indicates the TZ. A color version of this figure is available online.}
\label{Emissivity}
\end{figure}

The fraction of H$_2$ molecules at the level (1,3) as a function of the distance to central star is shown in Fig. \ref{Pov} for the standard PN. This fraction varies by a factor of 5, while the H$_2$ density varies more than 10 orders of magnitude (Fig. \ref{Density}). For all our models, the density of H$_2$ molecules in the level (1,3) and, as a consequence, the emissivity of the 1-0 S(1) line follow a radial profile similar to $n($H$_2)$. Figure \ref{Emissivity} shows the emissivity of the line H$_2$ 1-0 S(1) in the ionized region as a function of the distance to the central star for the standard PN model. The similarity of the radial profiles of the H$_2$ density and the 1-0 S(1) emissivity can be noticed by comparing Figs. \ref{Density} and \ref{Emissivity}. The emissivity of this line is more important in the TZ, making this zone the most important contributor to the total 1-0 S(1) line intensity of the ionized region.

According to our calculations, the level (1,3) is populated by both collisional and radiative mechanisms. The contribution of the chemical processes is negligible. For the standard PN, for example, collisions dominate the entire ionized region, while UV pumping is significant in the TZ. The relative importance of collisional over radiative processes for the level (1,3) population increases with increasing $T_{\star}$ and $n_H$, and also for decreasing $L_\star$, although this last effect is not very significant. Collisions dominate the excitation for models with $T_{\star} >$ 200\,000~K. Formation of H$_2$ on grain surfaces may have some effect on the H$_2$ energy level distribution, particularly for the case of higher rovibrational levels and high grain-to-gas ratio, but for the level (1,3) this effect is not important. Other processes of H$_2$ formation and destruction of H$_2$ do not contribute significantly to the energy level distribution of the molecule in the ionized region.

In addition to the 1-0 S(1) line emissivity, Fig. \ref{Emissivity} also shows the radial profiles of the emissivity of some atomic lines as calculated by our code. The lines are Br$\gamma$ in 2.166 $\mu$m, [\ion{N}{II}] $\lambda$6583, [\ion{O}{I}] $\lambda$6300, and [\ion{S}{II}] $\lambda\lambda $6716, 6731. These lines are mainly produced in the region where H is ionized. According to our models, the peak in the emissivity of the 1-0 S(1) line in the ionized region is very close to the region where the atomic lines [\ion{N}{II}], [\ion{O}{I}], and [\ion{S}{II}] are mainly produced. 

As mentioned in the Introduction, similarities between the morphology of images in the 1-0 S(1) line and those taken in the [\ion{N}{II}], [\ion{S}{II}] and [\ion{O}{I}] forbidden optical lines and hydrogen recombination lines (H$\beta$, H$\alpha$, and Br$\gamma$) have been noticed in some PNe. Since the H$_2$ 1-0 S(1) line can be produced in a more extended region than those lines, the similarity in the images is evidence that a neutral envelope is absent in those objects. In this case, all the H$_2$ emission is produced in the ionized region, and the molecule coexists with N$^+$, O$^0$, and H$^+$ in the TZ.

A correlation between the intensities of the 1-0 S(1) and [\ion{O}{I}] lines in PNe was found by \citet{Reay_etal_1988}. Although they associate this correlation with the existence of clumps inside the ionized region, it can also be naturally explained by both lines being produced in the same region, since the production of the [\ion{O}{I}] line is very effective in the TZ, as shown in Fig. \ref{Emissivity}.

\begin{figure*}
\centering
\includegraphics[width=14cm]{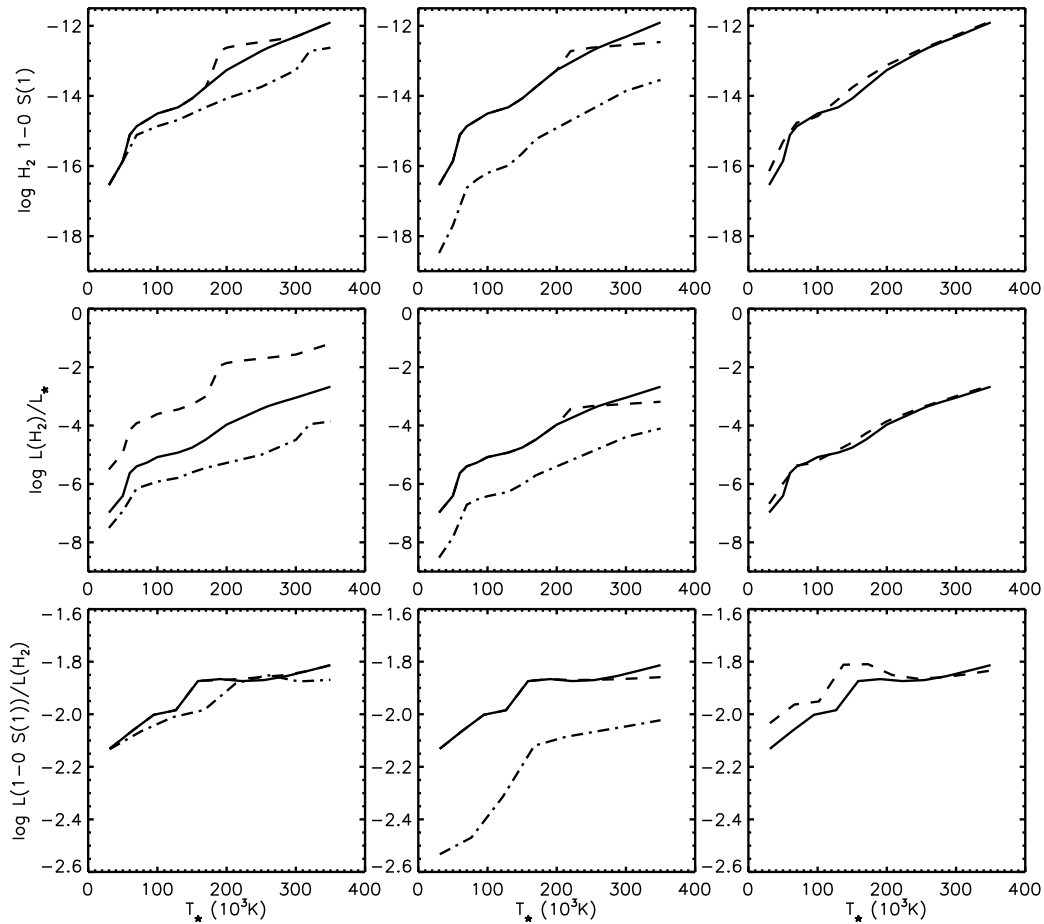}
\caption{Flux of the 1-0 S(1) line in arbitrary units (top row panels), ratio of the total H$_2$ IR lines luminosity to the central star luminosity (middle), and 1-0 S(1) luminosity to total H$_2$ IR lines luminosity ratio (bottom) as a function of $T_{\star}$. Curves for different $L_{\star}$, $n_\mathrm{H}$, and $M_\mathrm{d}/M_\mathrm{g}$ are shown, respectively, in the left, middle, and right panels in each row. The line styles are the same as in Fig. \ref{rm}. In the bottom left panel, the curves for $L_{\star}=$ 100 and 3000 $L_{\sun}$ coincide. The standard PN values are adopted for the parameters not mentioned.}
\label{Ints}
\end{figure*}

Since the 1-0 S(1) line is mostly produced in the TZ, the flux of this line is very sensitive to $T_{\star}$, as can be seen in the three top panels of Fig. \ref{Ints}, where the flux of the 1-0 S(1) line is shown as a function of $T_{\star}$. Curves for different $L_{\star}$ (left panels), $n_\mathrm{H}$ (middle), and $M_\mathrm{d}/M_\mathrm{g}$ (right) are shown. The 1-0 S(1) line flux does not depend much on $L_{\star}$ or $M_\mathrm{d}/M_\mathrm{g}$, but this flux decreases strongly if the model has a gas density higher than the standard PN value of 1000 cm$^{-3}$. The strong increase in the 1-0 S(1) flux with $T_{\star}$ we obtain with our models agrees with the results and conclusions from \citet{Phillips_2006}. After analyzing observational data from the literature, \citet{Phillips_2006} find a correlation between the detection of H$_2$ in PNe and their Zanstra temperatures, which he attributes to an effect of soft X-rays. The strong increase in the 1-0 S(1) flux with $T_{\star}$ can also explain why the detection of H$_2$ is more common in bipolar PNe \citep[Gatley's rule;][]{Zuckerman_Gatley_1988, Kastner_etal_1996}, since, according to \citet{Corradi_Schwarz_1995} and \citet{Phillips_2003}, these objects typically have higher $T_{\star}$. \citet{Stanghellini_2002} do not find a similar correlation in their analysis. This, however, may be due to their selection criterion that excludes PNe with high uncertainties in Zanstra temperature and, therefore, excludes high Zanstra temperature objects.

The ratio of the total luminosity in H$_2$ IR lines to the total luminosity of the central star is shown in the panels of the second row in Fig. \ref{Ints}. The sensitivity of this ratio with $T_{\star}$ is evident. PNe with smaller $L_{\star}$ or $n_\mathrm{H}$ also convert the stellar continuum in H$_2$ emission more efficiently. The change in the dust to gas ratio have a small effect in this ratio. Note that the total luminosity emitted by the H$_2$ lines may reach up to a few percent of the luminosity of the central star for models with a very hot and low-luminosity central star. This ratio is much lower for more typical PNe.

The 1-0 S(1) line intensity is around 1\% of the total H$_2$ IR emission in most models, as shown in the plots of the bottom row in Fig. \ref{Ints}. The ratio of the 1-0 S(1) line intensity to the total H$_2$ IR line emission does not depend much on $L_{\star}$ or $M_\mathrm{d}/M_\mathrm{g}$, but it increases for models with higher $T_{\star}$ and decreases for models with $n_\mathrm{H} >$ 10$^3$ cm$^{-3}$.

\begin{figure}
\centering
\resizebox{\hsize}{!}{\includegraphics{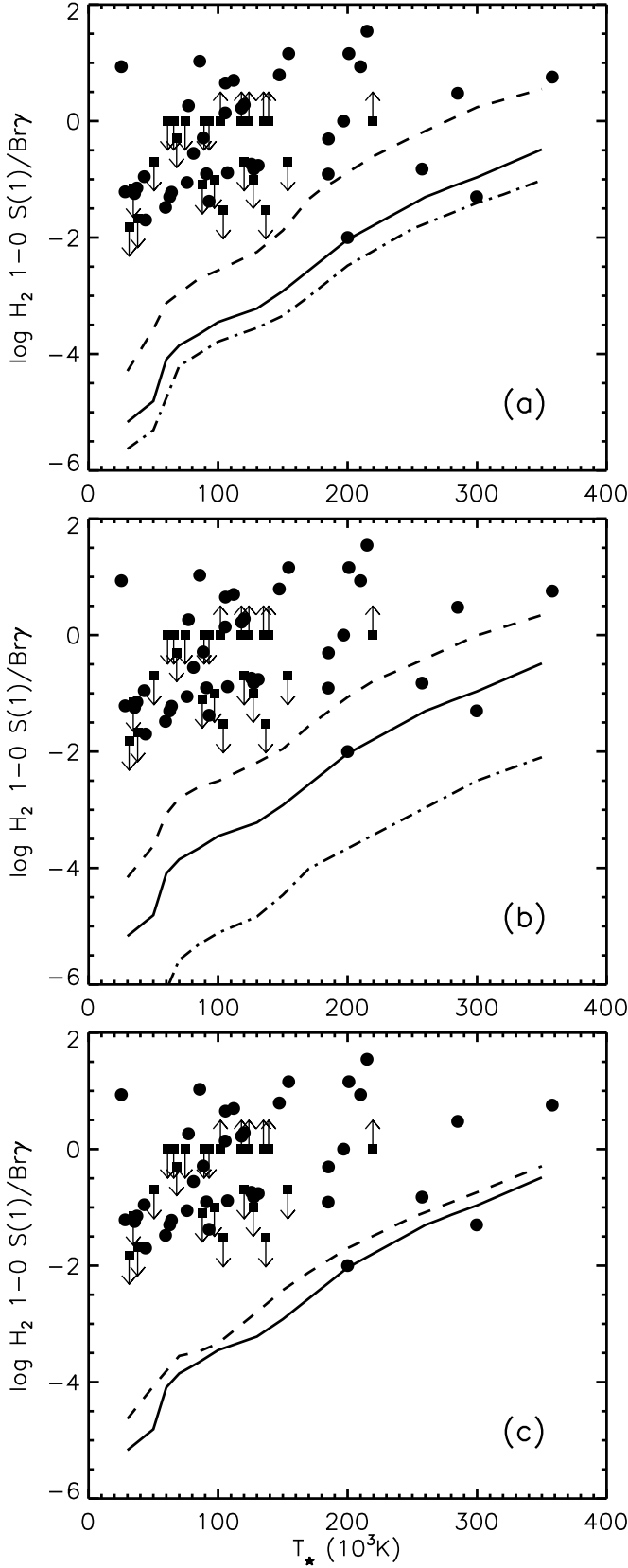}}
\caption{Line intensity ratio of H$_2$ 1-0 S(1) to Br$\gamma$ for the ionized region as a function of $T_{\star}$. Curves for different (a) $L_{\star}$, (b) $n_\mathrm{H}$, and (c) $M_\mathrm{d}/M_\mathrm{g}$ are shown. The line styles are the same as in Fig. \ref{rm}. Observed ratios are represented by dots. Boxes with up arrows are lower limits and with down arrows are upper limits. References for the observations are given in the text. The standard PN values are adopted for the parameters not mentioned.}
\label{H2toBrg}
\end{figure}

The ratio 1-0 S(1)/Br$\gamma$ in the ionized region as a function of the PNe model parameters is shown in Fig. \ref{H2toBrg}. Models with different $L_{\star}$, $n_\mathrm{H}$, and $M_\mathrm{d}/M_\mathrm{g}$ are shown in panels a, b, and c, respectively. Ratios obtained from observations are also shown. The observed values are from the following references: \citet{Beckwith_etal_1978}, \citet{Isaacman_1984}, \citet{Storey_1984}, \citet{Webster_etal_1988}, \citet{Aspin_etal_1993}, \citet{Allen_etal_1997}, \citet{Hora_etal_1999}, \citet{Davis_etal_2003}, \citet{Guerrero_etal_2000}, \citet{Rudy_etal_2001}, \citet{Kelly_Hrivnak_2005}, and \citet{Likkel_etal_2006}. In this figure we include ratios obtained from narrowband images and from long-slit spectra. In the latter case, the slit is often positioned here the H$_2$ emission is more intense, so the ratio 1-0 S(1) to Br$\gamma$ may be biased to higher values. In Fig. \ref{H2toBrg}, $T_\star$ represents the temperature of the stellar blackbody for models and the Zanstra or the energy-balance temperature for observations. For the Zanstra temperature, we used Tz(HeII) when available and Tz(HI) otherwise. In the few cases we did not find the Zanstra temperature, we adopted the value calculated through the energy-balance method \citep{PreiteMartinez_Pottasch_1983}. When more than one value (obtained by the same method) is found, an average is assumed. The temperatures were taken from the following references: \citet{Pottasch_etal_1978}, \citet{Martin_1981}, \citet{Kaler_1983}, \citet{Pottasch_1984}, \citet{Reay_etal_1984}, \citet{Shaw_Kaler_1985,Shaw_Kaler_1989}, \citet{FreitasPacheco_etal_1986}, \citet{Gathier_Pottasch_1988,Gathier_Pottasch_1989}, \citet{Gleizes_etal_1989}, \citet{Kaler_Jacoby_1989,Kaler_Jacoby_1991}, \citet{Jacoby_Kaler_1989}, \citet{PreiteMartinez_etal_1989,PreiteMartinez_etal_1991}, \citet{Walton_etal_1989}, \citet{Kaler_etal_1990}, \citet{Mendez_etal_1992}, \citet{Stasiska_etal_1997}, \citet{Bohigas_2001}, and \citet{Szyszka_etal_2009}.

Planetary nebulae can be radiation- or matter-bounded, depending on the amount of matter in the nebula and on the ionizing spectrum of the central star. In the first case, the ionizing spectrum cannot ionize the whole nebula. As a result, there is a neutral outer region. The emission of Br$\gamma$ is produced inside the H ionized region, while the emission of 1-0 S(1) line is produced in both the TZ and the neutral region. In our models, we assume that the nebula is radiation-bounded, therefore the line intensities are obtained by integrating the emissivity along the nebula, from the inner to the outer radius of the ionized region of the PN. Since in this work only the contribution of the ionized region is calculated, in the case of a radiation-bounded PN, only a lower limit is obtained for the H$_2$ line intensities and for the line ratio H$_2$ to Br$\gamma$. For a matter-bounded PN, the whole nebula is ionized, and our ratios are upper limits.

Figure \ref{H2toBrg} shows that the ionized region can be responsible for part or even the whole 1-0 S(1) emission of PNe. The ionized region can account for the whole emission for some PNe with high $T_\star$. In these cases, the hot star may be ionizing the whole (i.e., the PN is matter-bounded) or most of the nebula. 

Most PNe with detected H$_2$ emission, however, may have a molecular envelope, which could be responsible for the differences between the calculated and observed ratios. Contamination by other nearby H$_2$ lines is negligible according to our models. The spectral resolution in \citet{Hora_etal_1999}, from which most of the observation data were obtained, allows resolving the 1-0 S(1) line from a neighboring helium line, which may otherwise contaminate the molecular line.

\subsection{Other important H$_2$ lines}

The H$_2$ molecule emits a few thousand rovibrational lines in the IR. According to our models, the most intense lines are emitted in the range 1 to 29 $\mu$m, most of them with wavelengths from 1 to 5 $\mu$m (near IR). Above 5 $\mu$m the most intense lines are in general pure rotational lines, particularly S(0) to S(8) of the 0-0 band. The majority of the intense lines are produced by ortho levels, especially those from higher vibrational levels, since the ortho-para ratio of H$_2$ is normally higher than unity. Table \ref{table:1} lists those lines that are more intense than 1\% of Br$\gamma$ in at least 10\% of the models we ran, covering a variety of PN parameters. Assuming this criterion, the most intense lines in the ionized region are lines of the molecular bands 0-0, 1-0, 1-1, 2-0, 2-1, 3-1, 4-1, 4-2, and 5-3. Several of these intense lines have already been detected in PNe (they are indicated by an asterisk in Table \ref{table:1}). Several authors published observations of H$_2$ lines in the K band \citep[for example]{Ramsay_etal_1993, Hora_etal_1999, Vicini_etal_1999}. \citet{Hora_etal_1999} report the detection of more than 50 H$_2$ lines, with upper vibrational levels with quantum number up to 11, in the J, H, and K bands, which are rich in H$_2$ lines. They are atmospheric windows, which allows both ground and space-based observations. Molecular hydrogen lines were also detected in other spectral ranges in PNe. For instance, \citet{Matsuura_Zijlstra_2005} detected the lines S(1) to S(6) of the molecular band 0-0 in observations of \object{NGC 6302} with the Infrared Space Observatory (ISO); \citet{Bernard-Salas_Tielens_2005} published the detection of the lines 0-0 S(1) to S(7), 1-0 Q(1), 1-0 Q(3), 1-0 Q(5), 1-0 O(3), 1-0 O(4), 1-0 O(5), 1-0 O(6), and 2-1 O(3), also in observations made with ISO; the lines S(2) to S(7) of the 0-0 band were detected by \citet{Hora_etal_2006} in the Helix nebula (\object{NGC 7293}) with the Spitzer Space Telescope.

\begin{table}
\begin{minipage}[t]{\columnwidth}
\caption{Important H$_2$ lines of the ionized region of PNe}
\label{table:1}
\centering 
\renewcommand{\footnoterule}{} 
\begin{tabular}{l l l l l l}
\hline\hline
Line\footnote{Lines with (*) are already or possibly detected in PNe. By possibly detection we mean that a blend with or contamination by other line may prevent confirmation of the detection so far.} & $\lambda$($\mu$m) & Line & $\lambda$($\mu$m) & Line & $\lambda$($\mu$m) \\ 
\hline 
0.8286 & 4-1 S(7) &1.6750 & 1-0 S(19) &3.3718 & 0-0 S(23)\\
0.8306 & 4-1 S(9) &1.6877 & 1-0 S(9)* &3.3809 & 0-0 S(20)\\
0.8337 & 4-1 S(5) &1.7147 & 1-0 S(8)* &3.3876 & 0-0 S(24)\\
1.0519 & 2-0 S(10) &1.7480 & 1-0 S(7)* &3.4039 & 0-0 S(19)\\
1.0526 & 2-0 S(11) &1.7880 & 1-0 S(6)* &3.4379 & 2-1 O(5)\\
1.0536 & 2-0 S(9) &1.7904 & 2-1 S(9) &3.4384 & 0-0 S(18)\\
1.0576 & 2-0 S(8) &1.7963 & 2-1 S(19) &3.4856 & 0-0 S(17)\\
1.0608 & 2-0 S(13) &1.8358 & 1-0 S(5)* &3.5007 & 1-0 O(6)*\\
1.0641 & 2-0 S(7) &1.8528 & 2-1 S(7) &3.5470 & 0-0 S(16)\\
1.0733 & 2-0 S(6) &1.8920 & 1-0 S(4) &3.5926 & 1-1 S(21)\\
1.0851 & 2-0 S(5) &1.8947 & 2-1 S(6) &3.6198 & 1-1 S(19)\\
1.0998 & 2-0 S(4) &1.9449 & 2-1 S(5)* &3.6263 & 0-0 S(15)\\
1.1175 & 2-0 S(3) &1.9576 & 1-0 S(3)* &3.6522 & 1-1 S(18)\\
1.1204 & 3-1 S(9) &2.0041 & 2-1 S(4)* &3.6979 & 1-1 S(17)\\
1.1211 & 3-1 S(11) &2.0338 & 1-0 S(2)* &3.7245 & 0-0 S(14)\\
1.1240 & 3-1 S(8) &2.0656 & 3-2 S(5)* &3.7602 & 1-1 S(16)\\
1.1304 & 3-1 S(7) &2.0735 & 2-1 S(3)* &3.8075 & 1-0 O(7)\\
1.1320 & 3-1 S(13) &2.1218 & 1-0 S(1)* &3.8404 & 1-1 S(15)\\
1.1382 & 2-0 S(2) &2.1542 & 2-1 S(2)* &3.8464 & 0-0 S(13)\\
1.1397 & 3-1 S(6) &2.2014 & 3-2 S(3)* &3.8681 & 2-2 S(19)\\
1.1519 & 3-1 S(5) &2.2233 & 1-0 S(0)* &3.9391 & 2-2 S(17)\\
1.1622 & 2-0 S(1) &2.2477 & 2-1 S(1)* &3.9414 & 1-1 S(14)\\
1.1672 & 3-1 S(4)* &2.4066 & 1-0 Q(1)* &3.9968 & 0-0 S(12)\\
1.1857 & 3-1 S(3)* &2.4134 & 1-0 Q(2)* &4.0675 & 1-1 S(13)\\
1.1958 & 4-2 S(9) &2.4237 & 1-0 Q(3)* &4.0805 & 2-2 S(15)\\
1.1989 & 4-2 S(11) &2.4375 & 1-0 Q(4) &4.1622 & 1-0 O(8)\\
1.2047 & 4-2 S(7) &2.4548 & 1-0 Q(5)* &4.1810 & 0-0 S(11)\\
1.2263 & 4-2 S(5)* &2.4755 & 1-0 Q(6) &4.2236 & 1-1 S(12)\\
1.2330 & 3-1 S(1)* &2.5001 & 1-0 Q(7) &4.3138 & 2-2 S(13)\\
1.2383 & 2-0 Q(1)* &2.5278 & 1-0 Q(8) &4.4096 & 0-0 S(10)\\
1.2473 & 2-0 Q(3)* &2.5510 & 2-1 Q(1) &4.4171 & 1-1 S(11)\\
1.2616 & 4-2 S(3)* &2.5600 & 1-0 Q(9) &4.5757 & 1-0 O(9)\\
1.2636 & 2-0 Q(5) &2.5698 & 2-1 Q(3) &4.6563 & 1-1 S(10)\\
1.2873 & 2-0 Q(7)* &2.5954 & 1-0 Q(10) &4.6761 & 2-2 S(11)\\
1.2894 & 5-3 S(7) &2.6040 & 2-1 Q(5) &4.6947 & 0-0 S(9)\\
1.3107 & 5-3 S(5) &2.6269 & 1-0 O(2) &4.9533 & 1-1 S(9)\\
1.3116 & 4-2 S(1)* &2.6350 & 1-0 Q(11) &5.0529 & 0-0 S(8)\\
1.3188 & 2-0 Q(9) &2.6538 & 2-1 Q(7) &5.2390 & 2-2 S(9)\\
1.3240 & 3-1 Q(3) &2.6789 & 1-0 Q(12) &5.3304 & 1-1 S(8)\\
1.3420 & 3-1 Q(5) &2.7200 & 2-1 Q(9) &5.5115 & 0-0 S(7)\\
1.3472 & 5-3 S(3) &2.7269 & 1-0 Q(13) &5.8111 & 1-1 S(7)\\
1.3584 & 2-0 Q(11) &2.8025 & 1-0 O(3)* &6.1089 & 0-0 S(6)*\\
1.3684 & 3-1 Q(7) &2.8039 & 2-1 Q(11) &6.9091 & 0-0 S(5)*\\
1.4034 & 3-1 Q(9) &2.8361 & 1-0 Q(15) &7.2807 & 1-1 S(5)\\
1.4068 & 2-0 Q(13) &2.9061 & 2-1 Q(13) &8.0258 & 0-0 S(4)*\\
1.4295 & 4-2 Q(5) &2.9741 & 2-1 O(3)* &9.6649 & 0-0 S(3)*\\
1.4479 & 3-1 Q(11) &3.0039 & 1-0 O(4)* &12.2785 & 0-0 S(2)*\\
1.4592 & 4-2 Q(7) &3.2350 & 1-0 O(5)* &17.0346 & 0-0 S(1)*\\
1.6504 & 1-0 S(11) &3.3663 & 0-0 S(22) &28.2207 & 0-0 S(0)\\
1.6665 & 1-0 S(10) &3.3689 & 0-0 S(21) & &\\
\hline 
\end{tabular}
\end{minipage}
\end{table}

\begin{figure*}
\centering
\includegraphics[width=14cm]{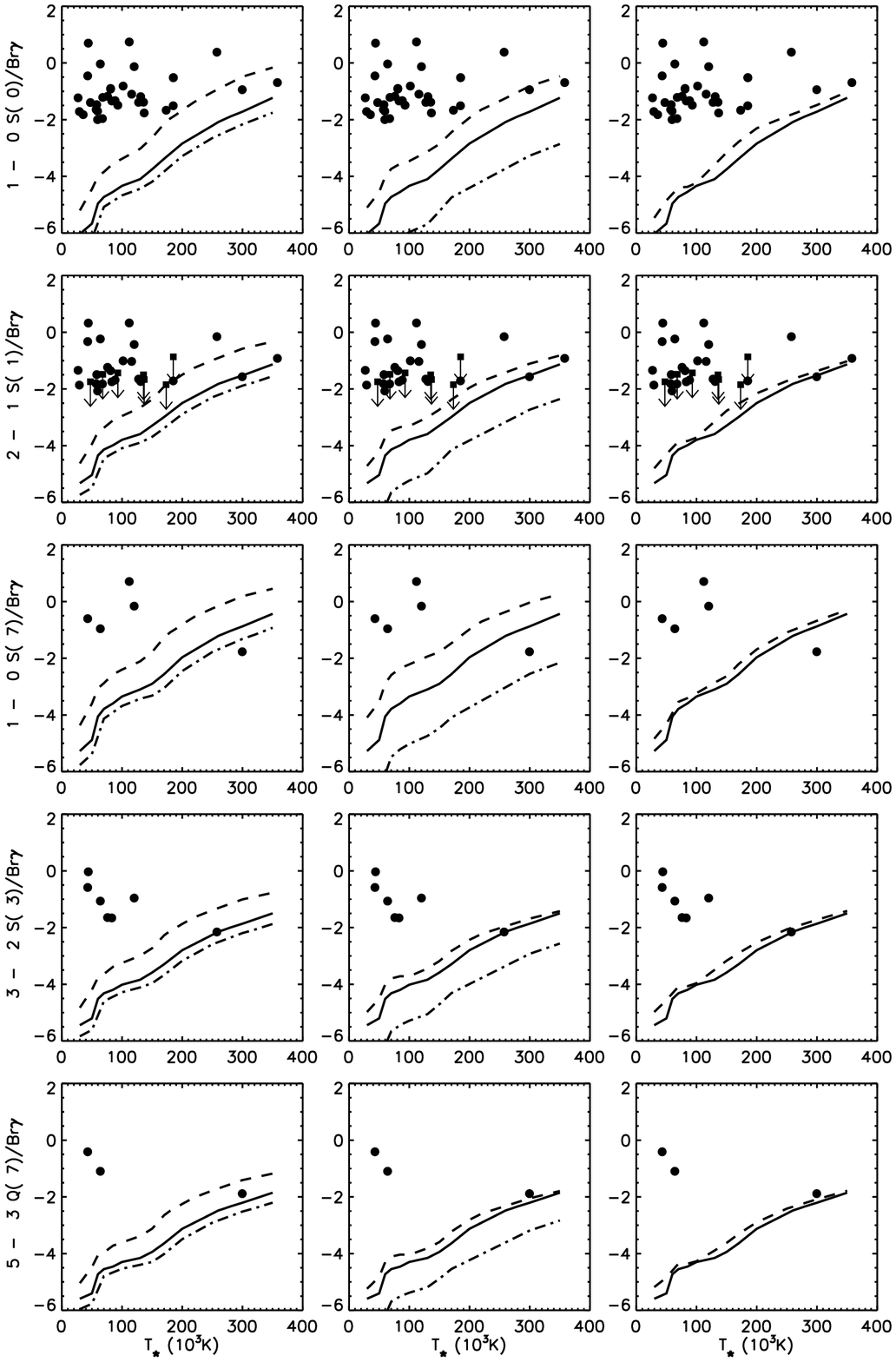}
\caption{Ratios of some H$_2$ lines to Br$\gamma$ for the ionized region as a function of $T_{\star}$. The models are the same as for Fig. \ref{rm}. The observed ratios are represented by dots. Boxes with down arrows are upper limits. The references for the observations are given in the text.}
\label{ratiolines}
\end{figure*}

Ratios of some H$_2$ lines to Br$\gamma$ are shown in Fig. \ref{ratiolines}. The plots are similar to those shown in Fig. \ref{H2toBrg}. As previously mentioned, in the ionized region, the production of the 1-0 S(1) line is more efficient in the TZ. Similarly, all H$_2$ lines are produced mostly in this zone. As a consequence, the intensity of the H$_2$ lines depends on the PN parameter in a similar same way as the 1-0 S(1) line, as can be seen in Fig. \ref{ratiolines}. 

Ratios obtained from observations of PNe are also included in Fig. \ref{ratiolines}. References for the central star temperature are the same as in Fig. \ref{H2toBrg}. The line ratios are obtained from \citet{Hora_etal_1999} and \citet{Sterling_etal_2008}. Our result show that the ionized region can account for part or even all the H$_2$ emission of these lines, particularly for PNe with hot central stars.

\subsection{Excitation mechanisms of the H$_2$ levels}

In all the analyzed models, collisional transitions are a major mechanism for population and depopulation of the H$_2$ energy levels, in the whole ionized region. This mechanism dominates the excitation of the lower vibrational levels. Collisions with ionized species are important in the inner (hotter and most ionized) zone of the models. The net effect of this mechanism is to populate levels with $J < 5$ by de-excitation of the upper levels (especially with $J > 7$). Collisions with neutral species, on the other hand, are more important in the TZ, where the temperature is moderate and neutral species are abundant. The net effect of this mechanism is to de-excitate the levels with lower rotational number, populating the upper rotational levels, especially those with lower vibrational numbers. The importance of collisions increase with $T_{\star}$, since the TZ is more extended in this case.

The rate of H$_2$ rovibrational radiative excitation is negligible. On the other hand, rovibrational de-excitation following electronic (UV pumping) or rovibrational collisional excitation is very important. The main effect of this mechanism is the excitation of H$_2$ from the lower rotational levels of $v = 0$ to the higher vibrational levels ($v > 3$) with $J \le 7$. For levels with $v \ge 9$ and $J \le 7$ UV pumping is the dominant mechanism. The contribution of the C$^-$ and C$^+$ states for the UV pumping rate is similar to the contribution of the B state. UV pumping is particularly important for H$_2$ excitation in the case of PNe with lower $n_\mathrm{H}$ and/or $T_\star$.

Formation pumping has a secondary effect on the H$_2$ population in the ionized region. Such an effect is only noticed for the upper levels ($v > 5$ and $J > 6$) in the outer fraction of PNe, especially for denser nebula or colder central stars. In this case, grain surface reaction and associative detachment are the main processes. Grain surface reaction may have a stronger effect for high dust density models.

The H$_2$ destruction processes do not affect the H$_2$ population in the ionized region. Photoionization has a small effect (less than 10\%) in populating of the lower vibrational levels in the inner zone of the nebula. However, since the H$_2$ density is very low in this region, it does not affect the intensities of the H$_2$ lines of PNe.

\section{Summary} \label{final}

In the present work we have studied the H$_2$ infrared emission from the ionized region of PNe. For this, we used the one-dimensional photoionization code Aangaba, in which we included the physics and chemistry of the H$_2$ molecule. This powerful tool can now be used to study the H$_2$ density, level population, line emission, processes of formation and destruction, and mechanisms of population and depopulation of the molecular energy levels as a function of the distance from the ionizing source of an ionized gaseous nebula.

Although there is observational evidence that at least part of the H$_2$ 1-0 S(1) line emission may originate inside the ionized region of PNe, the published studies of the molecular hydrogen emission of PNe do not usually take into account the contribution of the ionized region to this emission. One of the important conclusions of this work is that the H$_2$ emission of the ionized work can contribute significantly to the total emission observed in PNe, particularly for PNe with high $T_\star$. Comparison between the calculated and observed H$_2$ 1-0 S(1)/Br$\gamma$ ratio shows that the emission of the ionized region can be responsible for a substantial fraction of the total H$_2$ emission in such cases. Therefore it is important to use a code where the neutral and the ionized region are taken into account self-consistently.

In the ionized region, the H$_2$ IR emission lines are produced predominantly in the TZ, where the atomic lines [\ion{N}{II}], [\ion{O}{I}], and [\ion{S}{II}] are significantly produced. The partially ionized and warm gas in the TZ favors the formation and survival of H$_2$ molecules, as well as its IR line emission. The temperature of the central star is an important factor for the H$_2$ density and IR line intensity, since hotter stars produce more high-energy photons than colder stars. These photons can penetrate deep into the nebulae, producing the TZ. The 1-0 S(1) line intensity increases strongly with the increase in $T_{\star}$ in our models. This result agrees with the correlation between the detection of H$_2$ in PNe and their Zanstra temperatures found by \citet{Phillips_2006}. Furthermore, this can explain why the detection of this H$_2$ line is more common in bipolar PNe (Gatley's rule), given that these objects typically have higher $T_{\star}$. Although \citet{Reay_etal_1988} suggest that the correlation between the intensity of the 1-0 S(1) and [\ion{O}{I}] lines in PNe is due to the existence of clumps, such correlation can also be naturally explained by both lines being produced in the TZ.

The most intense H$_2$ lines are emitted in the range 1 to 29 $\mu$m, in the bands 0-0, 1-0, 1-1, 2-0, 2-1, 3-1, 4-1, 4-2, and 5-3. Several of these intense lines have already been detected in PNe. The 1-0 S(1) line is one of the more intense lines. The fraction of the 1-0 S(1) line to the total H$_2$ IR line emission is around 1\% (within a factor of less than 10) in all our models. 

Both collisions and UV pumping play important roles in the excitation of H$_2$ infrared lines in the ionized region. The effect of the excitation by UV pumping is important for levels with $v > 3$. The relative importance of collisions over UV pumping increases with the increase in $T_{\star}$ and $n_\mathrm{H}$. Grain surface reaction and associative detachment may only be significant for very high levels ($v > 5$ and $J > 6$), particularly in the cases of denser nebula or colder central stars. Grain surface reaction may be important for lower levels in the presence of large amounts of dust.

The effect of H$_2$ on the thermal equilibrium is insignificant in most models, except in the cases of a very high central star temperature or dust density, where the molecule cools the gas in the outer zone of the TZ, primarily through collisional excitation.

\begin{acknowledgements}

We are thankful to E. Roueff for providing the transition probabilities, the dissociation fractions, and the energies of the rovibrational levels of H$_2$, to D. Flower for providing collisional rate coefficients for the hydrogen molecule, and to A. Zijlstra for the careful reading of the paper. We also acknowledge the anonymous referee and the editor, M. Walmsley, for their suggestion for improving this paper. I.A. acknowledges the financial support of CAPES/Proex and FAPESP (998/14264-8).

\end{acknowledgements}

\bibliographystyle{H2inPNe}  
\bibliography{H2inPNe}

\end{document}